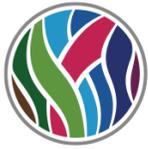

**IN PRESS**

# Arabidopsis plants perform arithmetic division to prevent starvation at night

Plants implement arithmetic division to optimize use of carbohydrate reserves and thus maintain metabolism and growth at night.

Antonio Scialdone (John Innes Centre), Sam Mugford (John Innes Centre), Doreen Feike (John Innes Centre), Alastair Skeffington (John Innes Centre), Philippa Borrill (John Innes Centre), Alexander Graf (ETH Zurich), Alison Smith (John Innes Centre), and Martin Howard (John Innes Centre)

**Abstract:**

Photosynthetic starch reserves that accumulate in Arabidopsis leaves during the day decrease approximately linearly with time at night to support metabolism and growth. We find that the rate of decrease is adjusted to accommodate variation in the time of onset of darkness and starch content, such that reserves last almost precisely until dawn. Generation of these dynamics therefore requires an arithmetic division computation between the starch content and expected time to dawn. We introduce two novel chemical kinetic models capable of implementing analog arithmetic division. Predictions from the models are successfully tested in plants perturbed by a night-time light period or by mutations in starch degradation pathways. Our experiments indicate which components of the starch degradation apparatus may be important for appropriate arithmetic division. Our results are potentially relevant for any biological system dependent on a food reserve for survival over a predictable time period.

http://dx.doi.org/10.7554/elife.00669

# Introduction

Organisms must control the rate of consumption of their stored food reserves to prevent starvation during periods when food acquisition is not possible. A classic example of this requirement is provided by the response to the light/dark cycle in plants. During the day, plants utilize solar energy for carbon assimilation through photosynthesis. During the night, when solar energy is unavailable, plants utilize stored carbohydrate - usually starch - to allow continued metabolism and growth. In the model plant *Arabidopsis thaliana* this phenomenon is essential for productivity: mutants with defects in either the accumulation or the degradation of starch have reduced productivity and exhibit symptoms of starvation (Usadel et al., 2008; Yazdanbakhsh et al., 2011). The leaf starch content of Arabidopsis increases approximately linearly with time during the day: more than half of the carbon assimilated via photosynthesis may be stored as semi-crystalline starch granules inside chloroplasts. At night, starch content decreases approximately linearly with time such that 95% of starch is utilized by dawn (Gibon et al., 2004; Graf et al., 2010). This pattern of utilization is extremely robust, and is achieved even when darkness comes unexpectedly early (Graf et al., 2010). It is also likely to be optimal for the efficient utilization of carbohydrate over the light/dark cycle (Gibon et al., 2004; Smith and Stitt, 2007; Graf and Smith, 2011; Stitt and Zeeman, 2012; Feugier and Satake, 2013). However, despite the high importance for plant productivity of precise control of starch degradation, the way in which such dynamics are generated is very poorly understood. One intriguing possibility is the existence of a mechanism that dynamically measures the starch content and the expected time to dawn, then arithmetically divides these two quantities to compute the appropriate starch degradation rate. Such a mechanism could ensure complete utilization of available starch reserves at dawn despite variation in both the starch content at the onset of darkness and the subsequent duration of darkness. Consistent with this idea, we have recently shown that computation of the appropriate starch degradation rate in a normal night requires the circadian clock (Graf et al., 2010), which indicates how information about the expected time to dawn is obtained.

Although levels of transcripts for starch-degrading enzymes undergo large daily changes, levels of the enzymes themselves do not (Smith et al., 2004; Lu et al., 2005; Yu et al., 2005; Kötting et al., 2005; and our unpublished data). Therefore, it is likely that an arithmetic division mechanism would control flux through starch degradation at a post-translational level. Post-translational control would also permit swifter modulation of the catalytic capacity of these abundant proteins than would be possible via transcription/translation. Accordingly, in this paper we propose appropriate analog division mechanisms that operate through post-translational chemical kinetics. More generally, we also consider analog chemical kinetic schemes that generate addition, subtraction and multiplication operations. To the best of our knowledge the starch degradation system constitutes the first concrete realization of such arithmetic operations in biology. In this context, we therefore introduce two mathematical models which can both implement arithmetic division between the starch content and the expected time to dawn. We then successfully test predictions from the two models by examining the pattern of starch degradation in abnormal light/dark cycles and in a range of mutant plants defective in components of the starch degradation apparatus. Finally, our experiments also indicate which components of the starch degradation apparatus may be important for the appropriate implementation of arithmetic division.

# Results

We first investigated the robustness of a potential arithmetic division calculation to perturbations in both the numerator (starch content) and denominator (expected time to dawn). Previously, we showed that the rate of starch degradation is appropriately adjusted in response to an unexpectedly early night (imposition of darkness 8 h after dawn on plants grown in 12 h light, 12 h dark cycles) (Graf et al., 2010). We found that adjustment also occurs in response to an unexpectedly late night. Plants grown in 12 h light, 12 h dark cycles were subjected to darkness at either 12 h or 16 h after dawn. In both cases starch content decreased approximately linearly with time during the night, but with different slopes such that starch reserves were almost exhausted by dawn (Figure 1A). We also found that similar adjustments could be performed in a *cca1/lhy* circadian clock mutant which has a free-running period of less than 24 h (Alabadi et al., 2001). In 12 h light, 12 h dark cycles, this mutant degrades its starch by approximately 21 h after dawn, rather than the normal 24 h (Graf et al., 2010, Figure 1B).When subjected to an unexpected early night, the starch degradation rate in the mutant was adjusted, such that starch reserves were again exhausted at around 21 h after dawn (Figure 1B). Appropriate adjustments of the rate of starch degradation also occurred in wild-type plants in which environmental manipulations produced different starch contents at the end of the 12 h light period. A subset of a uniform batch of plants was transferred to a reduced light level for a single light period, leading to a two-fold reduction in the starch content at the onset of darkness. For these and control plants subjected to normal light levels, starch content decreased approximately linearly with time during the night, but with different slopes such that starch reserves were almost exhausted by dawn in both cases (Figure 1C). Appropriate adjustment of starch degradation also occurred in subsets of plants exposed to three different regimes of varying light intensity over a single light period that generated two different starch contents at the onset of darkness (Figure 1 – Figure Supplement 1). To investigate whether this phenomenon is widespread among plants, we examined the model grass *Brachypodium distachyon.* Ancestors of Arabidopsis and *Brachypodium* diverged at least 140 Myr ago. Starch content in *Brachypodium* leaves increased through the light period. As in Arabidopsis, the approximately linear decrease of starch with time following either a normal night (12 h after dawn) or an unexpectedly early night (8 h after dawn) was such that starch reserves were almost depleted by dawn (Figure 1D).

Overall, these results demonstrate that the control of starch degradation at night to achieve almost complete consumption at the expected time of dawn can accommodate unexpected variation in the time of onset of darkness, starch content at the start of the night, and patterns of starch accumulation during the preceding day. Although the rate of degradation is different in a circadian clock mutant with an altered period from that in the wild-type, the capacity to adjust starch degradation in response to an unexpectedly early night is not compromised. It is also likely that the mechanism underlying this control is present in evolutionarily-distant species of plants.

Robust computation of the appropriate starch degradation rate in response to perturbations of both the expected time to dawn and starch content is clearly consistent with the implementation of arithmetic division. Since it is conceptually unclear how such a computation might be performed, we turned to mathematical modeling to generate possible mechanisms.

The fact that starch exists as large polymers that are mostly inaccessible within granule matrices means that a separate measure is likely required to provide information about the total amount of starch present at any given time. Hence, we assume the existence of a soluble molecule $S$ whose concentration is proportional to the amount of starch in a granule. Since plants are able to adjust the rate of starch degradation according to variations in two independent quantities (the expected time to dawn and the amount of starch present), two separate species of molecule are clearly required. Therefore, we further assume the existence of a molecule $T$ whose concentration encodes information about the expected time to dawn.

In our first model the $T$ molecule concentration is proportional to the expected time to dawn, except during a period after dawn when its value must be reset (Figure 2B).To compute the appropriate degradation rate, the $S$ and $T$ molecule concentrations must therefore be arithmetically divided. We propose that computations of this form can be carried out most simply using analog chemical kinetics. As shown in Figure 2A, it is straightforward to perform addition, subtraction and multiplication operations. Subtraction can be implemented through efficient sequestration and multiplication by a two-species chemical reaction. Division is slightly more intricate, but can be implemented using the model introduced in Figure 2D (other conceptually similar models are discussed in the Materials and Methods). Here, $S$ molecules associate with the starch granule surface, where they permit starch degradation (presumably in combination with other elements, as shown in Figure 2D) and where the $S$ molecules can also be degraded. $T$ molecules inhibit $S$ molecule and starch degradation by binding to $S$ and causing its detachment from the granule surface. In this way, it can be seen intuitively that a division-like operation can be implemented (for rigorous calculations, see Materials and Methods). In Figure 2F,H,J, we show the best fit of this model to the data from Figure 1A,B,C with good results.

A second distinct possibility also exists for computing the appropriate degradation rate. We now assume that the $T$ molecule concentration increases as the expected time of dawn approaches, before being reset. If this increase is such that the $T$ molecule concentration is approximately proportional to 1/(expected time to dawn) (Figure 2C) then the appropriate degradation rate can be computed by multiplying the $S$ and $T$ concentrations. This is implemented by the reaction scheme shown in Figure 2E: $S$ molecules associate with the starch granule surface, where they recruit $T$ molecules from the stroma. The resulting molecular complex permits the degradation of starch (again presumably in combination with other elements, as shown in Figure 2E) and of the $S$ molecule itself.   The output of this second model is very similar to that of the first model (fits to data of Figure 1A,B,C shown in Figure 2G,I,K) in that a division computation is still performed, but now the timing information is encoded differently in the $T$ molecule concentration. One potential difficulty is the need to pre-compute the reciprocal of the expected time to dawn. A simple possible scheme to achieve this goal is outlined in the Materials and Methods. Of course, a combination of the two above models is also possible, involving both multiplication and division by factors dependent on the expected time to dawn, such that overall an appropriate division computation is still performed. However, the additional complexity required for such implementations makes such a combined model less likely.

Importantly we assume that the granule surface area does not limit the reaction rate as the granule shrinks, consistent with an approximately linear decrease of starch content with time. Accordingly, in both models the degradation reactions (for both the starch and the $S$ molecules) occur only in region(s) of granule surface of overall fixed area as each granule shrinks. This would be the case if one or more of the additional components required for starch degradation (illustrated in Figure 2D,E) is present at a fixed number on the granule surface as the granule shrinks. Clearly the assumption of a non-limiting surface area cannot remain true if the granule shrinks to very small volumes, as could happen at the end of the night. However, our models still fit the experimental data well even at these times (see Figure 2F-K).

Taken together, our modeling results show that arithmetic division (as well as other arithmetic operations) can be implemented simply using analog post-translational chemical kinetics. Furthermore, output from both models fits our experimental data well. We note that there is some variation in the fitted parameters for both models (see Tables 3-6), which arises due to variation between experimental data sets even for a single genotype. Such variation is widely observed for measurements of primary metabolites over the light/dark cycle and could arise from batch to batch variation in expression levels of degradation or clock components.

As the fit of the two models to the data is equally good, distinguishing between them is currently challenging. However, we can test two critical predictions common to both models. The first prediction is that the degradation rate is continuously computed via arithmetic division during the night. Such a scheme clearly has advantages in its flexibility and potential to recover from unexpected perturbations. To test whether this prediction is correct, we interrupted a normal night with a period of light, ending 5 h before the expected time of dawn. During this night-time light period starch accumulated to levels very similar to those present at the end of the day-time light period. Comparison of the rate of starch degradation following the night-time light period with the rate before this period, and with the rate that would have been expected over a normal 12-h night, allowed a robust assessment of whether the degradation rate had been reset. We confirmed that the night-time light period did not re-entrain the circadian clock, by monitoring expression of the clock gene *LHY* (see Materials and Methods and Figure 3 – Figure Supplement 1). For three independent experiments we found that the rates of starch degradation immediately following the night-time light period (between 19 h and 21 h after dawn) were significantly greater than both the corresponding actual rates before this period (between 12 h and 14 h after dawn) and the corresponding rates that would have been expected during a normal 12-h night, see insets in Figure 3A,B,C (p-values = $1.6 \times 10^{-3}$ and $3.8 \times 10^{-5}$ respectively, details of the statistical analysis in the Materials and Methods). The latter rates are those that would have ensured the complete depletion of the starch content measured at 12 h at the expected time of dawn (24 h). We obtained the same result by changing the duration and the starting time of the night-time light period (Figure 3D). This increase in starch degradation rate is consistent with our prediction that the rate is continuously computed during the night, rather than set only once at the first onset of darkness. In Figure 3 we show the best fits of the first model to the data, with good results. As before, the second model produced very similar fits (see Figure 3 – Figure Supplement 2).

The second prediction from the models concerns the effects on starch degradation of perturbations in components of the arithmetic division or degradation apparatus. To reproduce the normal pattern – in which starch is almost completely degraded by the expected time of dawn – our models require that a factor involved in specifying the relative degradation rates of the starch and *S* molecules must be fine-tuned to one (see Materials and Methods). If instead this factor is less than one, so that the starch is degraded too slowly, then rather than complete degradation the models predict that only a certain percentage of the starch will be consumed during the night regardless of the starch content at the end of the preceding light period (for full calculations, see Materials and Methods). Accordingly, we predict that perturbations to parts of the arithmetic division or degradation apparatus might result in degradation of only a fraction of starch during the night. Interestingly, several mutants defective in proteins involved in, or related to, starch degradation exhibit approximately linearly decreasing starch content with time during the night, but have an elevated starch content at the end of the night. These include beta-amylase mutants *bam3* and *bam4*, the debranching enzyme mutant *isa3*, and mutants lacking phosphoglucan water dikinase (*pwd*, also called *gwd3*), a glucan phosphate phosphatase (*sex4*) and a glucan phosphate phosphatase-like protein (*lsf1*) (Smith, 2012). We showed previously that this abnormal pattern of starch turnover was rapidly regained in *lsf1* mutants that were transferred to normal light/dark cycles after starch was reduced to very low levels by prolonged darkness (Comparot-Moss et al., 2010). We re-analyzed these data and also performed a similar experiment on the *sex4* mutant. For both mutants, we found that the fraction of end-of-light period starch content degraded during the night was approximately the same on successive nights (around 30% for the *sex4* mutant and 45% for the *lsf1* mutant), regardless of the starch content at the end of the respective preceding light period. This resulted in a progressive increase in the end-of-night starch content to an approximately constant value over three to four days after return to normal light/dark cycles (Figure 4A,B). Thus the pattern of starch degradation in these mutants is as predicted from the models for a situation in which the above factor is incorrectly set to a value less than one.

We next attempted to gain experimental insight into how the arithmetic division mechanism modulates flux through the starch degradation pathway. In our models the computation of the starch degradation rate is the result of interactions between the molecules encoding the information about starch content and time to dawn ($S$ and $T$) and the starch degradation apparatus. If this idea is correct, we expect that mutants lacking components of the degradation apparatus involved in these interactions may also lack the ability to adjust the rate of starch degradation in response to variation in starch content or time of onset of darkness.

To look for a mutant in this category, we imposed an unexpectedly early night on six mutants, each lacking a protein involved in starch degradation (*lsf1, sex4, bam3, bam4, isa3, pwd*). All six mutants show approximately linear decreases in starch content with time during the night, but they have higher end-of-night starch contents than wild-type plants. This mutant collection covers the majority of currently-known components of the chloroplastic starch degradation apparatus. For five of the six mutants, the rates of starch degradation (as determined from linear fits) were lower during an unexpectedly early night (starting 8 h after dawn) than in a normal night (starting 12 h after dawn), as is the case in wild-type plants (Figure 4C,D and Figure 4 – Figure Supplement 1). To quantify the adjustment of the starch degradation rates in the mutants, we calculated the ratio $R$ between the degradation rates (normalized by the respective end-of-light period starch content) during the normal and the early night. For wild-type plants, the expected value of $R$ is $16/12 \approx 1.33$, i.e., the ratio between the expected lengths of the early and the normal night. In Figure 4F we show that the values of $R$ for five of the six mutants are consistent with the wild-type value.

However for the sixth mutant, *pwd*, the rate of starch degradation was not adjusted in response to an unexpectedly early night (Figure 4E,F), as we found an $R$ which was significantly different from the wild-type value (p-value = 0.01, details of the statistical analysis in the Materials and Methods). This finding indicates that PWD - phosphoglucan, water dikinase - may be a node at which information about expected time to dawn and starch content is integrated to set an appropriate rate of starch degradation.

PWD contributes to a cycle of phosphorylation and dephosphorylation of glucosyl residues within starch polymers that is essential for normal starch degradation. Two enzymes – glucan water dikinase (GWD) and PWD – add phosphate groups to starch polymers at the granule surface, and two further enzymes SEX4 and LSF2 – remove them. All four enzymes are necessary for normal rates of starch degradation, and loss of GWD almost completely prevents degradation at night (Smith, 2012). The phosphate groups are thought to disrupt the ordered packing of the starch molecules at the granule surface, allowing access to starch degrading enzymes (beta-amylases and isoamylase 3). Subsequent removal of the phosphate groups is essential for full degradation of the starch polymers because the phosphates block the action of the exo-acting beta-amylases (Baunsgaard et al., 2005; Kötting et al., 2005; Ritte et al., 2006; Hejazi et al., 2009). The cycle as a whole is thus an attractive candidate for integration of factors that modulate flux through starch degradation.

Because the phosphorylation/dephosphorylation cycle modifies the granule surface, it is a potential candidate not only for flux modulation but also for the storage of information about starch content. To discover whether phosphate groups may provide information about starch content, we measured the amount of granule-bound phosphate (on the 6-position of glucosyl residues, representing about 80% of the total phosphate) over the light/dark cycle. If phosphorylation simply tracks starch polymer synthesis during the day, phosphate levels per unit mass of starch will be constant and will thus contain no information about starch content. Surprisingly, however, we found a large increase/decrease in the level of phosphate per unit mass of starch over the light/dark cycle (Figure 5). This discovery implies that the $S$ molecule may be a modulator of activities of enzymes of the phosphorylation/dephosphorylation cycle, thus generating a daily pattern of change in the accessibility of the granule surface to hydrolytic enzymes that approximately tracks starch content.

## Discussion

Overall, our work provides a new framework and perspective for understanding the control of reserve utilization in plants at night. Our experiments provide strong support for an implementation of arithmetic division in night-time starch degradation. We used mathematical modeling to generate two simple mechanisms capable of analog implementation of such an operation. Predictions from the models were then verified and a potential point identified at which the division computation could be integrated into the starch degradation pathway. We also showed that the phosphorylation state of the starch granule surface could provide information about starch content through the day.

Our analysis may also be relevant to a broader class of biological processes, where food reserves accumulated in advance of periods of predictable length without further food intake are just sufficient to permit survival to the expected end of the period. For example, migrating little stints (*Calidris minuta*) arriving at their Arctic summer breeding grounds after a 5000 km journey have sufficient remaining lipid reserves for an average of only 0.6 days (Tulp et al., 2009). During the four-month fast period of egg-incubating male emperor penguins (*Aptenodytes forsteri*), lipid reserves are used such that they reach a critical depletion level at approximately the point at which the females are expected to return. Unexpected extension of the fast period leads to catabolism of protein and abandonment of offspring in favor of hunting for food (Groscolas and Robin, 2001). As in Arabidopsis leaves, the rate of reserve utilization in these examples can potentially be computed by arithmetically dividing the reserve levels by the anticipated time of fasting.

It is a longstanding idea that cells are able to use proteins to store and process information through networks of interactions (Bray, 1995). Understanding how such biochemical networks work and what kind of computations they perform is an ongoing challenge (see Deckard and Sauro, 2004, Lim et al., 2013). Our analysis here has underlined the utility of analog chemical kinetics in performing arithmetic computations in biology. Importantly, we have for the first time provided a concrete example of a biological system where such a computation is of fundamental importance. This contrasts to previous work where elegant theoretical implementations of arithmetic operations lacked specific biological applications (Cory and Perkins, 2008, Buisman et al., 2009).  Analog chemical kinetic approaches may potentially also be useful for calculations in synthetic biology applications (Benenson, 2012), where they are likely to prove much simpler to implement than alternative schemes based on much more complex digital circuitry.

# Materials and Methods

## Plant material
*Arabidopsis thaliana* (in the Col0 background, except for *cca1/lhy* and its wild-type which were in the Ws background) and *Brachypodium distachyon* (Bd21) plants were grown as in Graf et al., 2010 on soil in 12 h light, 12 h dark with 200 $\mu$mol quanta m$^{-2}$ s$^{-1}$ illumination at a constant temperature of 20 ºC for 21 days.

## Measurement of starch
Plants were harvested and extracted in dilute perchloric acid for analysis of starch, which was then quantified enzymatically as previously described (Graf et al., 2010).

## Transcript analysis
RNA was extracted from plant material and qPCR was performed as in Graf et al., 2010. Oligonucleotide primer sequences were as follows:

| Primer | Sequence 5' to 3' |
|---|---|
| LHY-F AT1G01060 | GACTCAAACACTGCCCAGAAGA |
| LHY-RAT1G01060 | CGTCACTCCCTGAAGGTGTATTT |
| ACT2-F AT3G18780 | ACTTTCATCAGCCGTTTTGA |
| ACT2-R AT3G18780 | ACGATTGGTTGAATATCATCAG |

## Starch phosphate measurements
Starch granules were prepared as in (Ritte et al., 2000) and used immediately without drying. Granules were resuspended in two pellet volumes of water and boiled for 15 min then digested with 20 U amylogucosidase (Roche) and 2 U $\alpha$-amylase (Megazyme) in 100 mM Na acetate pH 4.8 for 9 h at 37°C. Glucose was assayed enzymatically following (Hargreaves and ap Rees, 1988) and glucose 6-phosphate was measured enzymatically using the fluorimetric assay of (Zhu et al., 2009).

# Mathematical Modelling

## Quantification of the starch content and the expected time to dawn

Starch is laid down as semi-crystalline granules in chloroplasts in the light. Granule surfaces are then subject to degradation during periods of darkness. We assume here that the computation of the degradation rate is performed autonomously inside each chloroplast. In order to correctly compute the appropriate starch degradation rate, knowledge of the total chloroplast starch content is required. We assume that there exists a molecule $S$ whose number $N_S^{tot}$ is proportional to the total starch content $\Delta S^{tot}$ in a chloroplast at the end of a light period:

$$\Delta S^{tot} = \alpha N_S^{tot} .$$

We also assume that the $S$ molecule can exchange rapidly between the granule surfaces and the surrounding chloroplast compartment. Moreover, we propose the existence of a molecule $T$ that encodes information about the expected time to dawn. The $T$ molecule dynamics will presumably be controlled by the circadian clock and are assumed to be independent of the starch degradation process.

In the first model (Figure 2B and 2D), the $T$ concentration in the compartment surrounding the granules, [$T_C$], is proportional to the expected time to dawn $\Delta t = t_{day} - t$ during most of the light/dark cycle. Note, however, that this cannot be true at all times, as around the time of expected dawn, after [$T_C$] has dropped to low levels, its value must be reset. So, if dawn occurs at $t=0$, we assume that:

**Eq. 1**

$$[T_C] = \begin{cases} \beta \dfrac{t_{day} - t_r}{t_r} t & 0 \le t \le t_r \\ \beta (t_{day} - t) & t_r \le t \le t_{day} \end{cases} ,$$

where $\beta$ is a constant, $t_{day}$ is the period of the light/dark cycle and $t_r$ is the reset-time after which [$T_C$] starts tracking the time to expected dawn (see Figure 2B).

These dynamics could be the result of production of $T$ at a constant rate, with a comparatively low degradation rate for $0 \le t \le t_r$. During $t_r \le t \le t_{day}$, [$T_C$] is assumed to decrease linearly with time, with a rate $\beta$. This could happen through an efficient sequestration of $T_C$ by another molecule which is produced at a rate $\beta$ and is degraded at a comparatively low rate for $t_r \le t \le t_{day}$. These simple [$T_C$] dynamics can reproduce the available experimental data on starch degradation during the hours after dawn. Experimentally, if the onset of darkness unexpectedly occurs only 6 h after the previous dawn in plants previously grown in 12 h light, 12 h dark conditions, the starch is degraded too quickly, while for an unexpected onset of darkness at 8 h, the starch degradation occurs at the appropriate rate (Lu et al., 2005; Graf et al., 2010). This can be explained by the above model, with the $T$ dynamics given in Eq. 1 and assuming a reset time $t_r \gtrsim 8$ h, as we have [$T_C$]< $\beta\Delta t$ during the resetting period, and hence the starch degradation rate will initially be too high.

Nevertheless, if the early night condition is maintained each day for a few days in succession, acclimation occurs: the appropriate starch degradation rate is computed even after a light period as short as 4 h. This suggests that $t_r$ can be reduced, thereby resetting [$T_C$] more quickly, and hence allowing the arithmetic division mechanism to work effectively under the new light/dark cycle conditions.

In the second model (Figure 2C and 2E), we assume that for most of the light/dark cycle [$T_C$] is proportional to $1/(t_{day}-t)$. Again, this cannot be true around dawn, where [$T_C$] must be reset. For the sake of simplicity, we assume that during the reset period [$T_C$] has a linear profile with time:

$$[T_C] = \begin{cases} \dfrac{\beta}{t_{day} + t_{r1} - t_{r2}} \left( \dfrac{t + t_{day} - t_{r2}}{t_{day} - t_{r1}} + \dfrac{t_{r1} - t}{t_{day} - t_{r2}} \right) & 0 \le t \le t_{r1} \\[2ex] \dfrac{\beta}{t_{day} - t} & t_{r1} \le t \le t_{r2} \\[2ex] \dfrac{\beta}{t_{day} + t_{r1} - t_{r2}} \left( \dfrac{t - t_{r2}}{t_{day} - t_{r1}} + \dfrac{t_{day} + t_{r1} - t}{t_{day} - t_{r2}} \right) & t_{r2} \le t \le t_{day} \end{cases} ,$$

where $\beta$ is a constant, and $t_{r1}$ and $t_{r2}$ define the time interval in which $[T_C]$ is proportional to $1/\Delta t$ (see Figure 2C). In the second model, these dynamics can also explain why starch degradation is too rapid following a very early and unexpected onset of darkness. During the resetting period after dawn $[T_C] > \beta/\Delta t$, and therefore the starch degradation rate will initially be too high.

One apparent difficulty with the second model is in finding a simple way to generate a $T$ molecule concentration that scales as $1/\Delta t$. In fact, kinetics of this kind are simple to produce. For example, if $T$ promotes further production of the $T$ molecule itself, with a quadratic nonlinearity, then, away from saturation, we have $\dfrac{d[T_C]}{dt} = \eta[T_C]^2$. It is straightforward to show that the solution to this equation is:

$$[T_C](t) = \dfrac{[T_C](t_{r1})}{1 - \eta(t - t_{r1})[T_C](t_{r1})}.$$

For $[T_C](t) = \beta/\Delta t$ as required, we then find $[T_C](t_{r1}) = \beta/(t_{day} - t_{r1})$ and $\eta = 1/\beta$.

Of course, more complex schemes could also generate a similar behavior; we simply wish to point out here that a $1/\Delta t$ behavior for $[T_C]$ is a plausible assumption.

## Two models that appropriately calculate the starch degradation rate

We now describe in full mathematical detail the models introduced in the main text.

### Model 1

In this model, division is implemented by dividing the $S$ concentration by the $T$ concentration, where the $T$ molecule concentration is proportional to $\Delta t$ for most of the light/dark cycle. Such a division can be implemented by the following set of reactions:

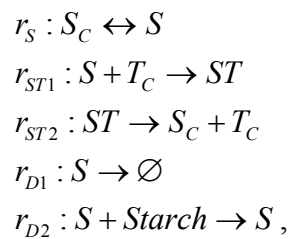

$r_S : S_C \leftrightarrow S$

$r_{ST1} : S + T_C \rightarrow ST$

$r_{ST2} : ST \rightarrow S_C + T_C$

$r_{D1} : S \rightarrow \varnothing$

$r_{D2} : S + Starch \rightarrow S ,$

where the subscript "$C$" refers to molecules in the compartment surrounding the granules. The reversible $r_S$ reaction describes exchange of the $S$ molecules between the granule surface and surrounding compartment, with forward reaction parameter $f_S$ and backward rate $b_S$. The $T$ molecules can be recruited by the $S$ molecules on the granule surface (reaction $r_{ST1}$, reaction parameter $f_{ST1}$) and the resulting complex can then dissociate leading to the detachment of $S$ and $T$ from the granule surface (reaction $r_{ST2}$, rate $f_{ST2}$). The $S$ molecule on the granule can be degraded (reaction $r_{D1}$, reaction parameter $f_{D1}$) or it can permit starch degradation (reaction $r_{D2}$, reaction parameter $f_{D2}$).

In all the calculations in this section, we assume that $b_S \gg f_{D1}$ and that the dynamics of the $S$ and $T$ molecules can be taken to be in quasi-steady-state. As described in the main text, we assume that the degradation reactions (for both the starch and the $S$ molecule) can only occur in a region of overall fixed area $A_d$ as each granule shrinks. We also assume that all granules in a given chloroplast are approximately equal in area and volume. The differential equations describing the dynamics of the total number of $S$ molecules, $N_S^{tot}$, and the total amount of starch in a chloroplast, $\Delta S^{tot}$, are:

**Eq. 4**

$$\frac{dN_S^{tot}}{dt} = n\frac{dN_S}{dt} = -nf_{D1}[S]A_d$$

$$\frac{d\Delta S^{tot}}{dt} = n\frac{d\Delta S}{dt} = -nm_S f_{D2}[S]A_d = -n\mu = -\mu^{tot},$$

where $m_S$ is the amount of starch degraded by an $S$ molecule in a single degradation reaction, $n$ is the number of granules in the chloroplast, $\mu$ is the starch degradation rate for an individual granule and $\mu^{tot}$ is the total starch degradation rate.

Using Eq. 4, we find that $\frac{d\Delta S^{tot}}{dt} = -\mu^{tot} = \frac{m_S f_{D2}}{f_{D1}}\frac{dN_S^{tot}}{dt}$. We assumed that at the beginning of the dark period $\Delta S^{tot} = \alpha N_S^{tot}$. It is easy to show that, if $\alpha = m_S \frac{f_{D2}}{f_{D1}}$, such a proportionality between the starch content and the number of $S$ molecules continues to hold true throughout the degradation process. The total number of $S$ molecules is $N_S^{tot} = [S_C]V + n[S]A + n[ST]A$, where $V$ is the compartmental volume surrounding the granules, and $A$ is the surface area of a granule. Using the above quasi-steady state assumption and the law of mass action, we obtain:

**Eq. 5**

$$[S] = \frac{k_S[S_C]}{1 + k_{ST1}[T_C]}, \quad [ST] = \frac{f_{ST1}}{f_{ST2}}\frac{k_S[S_C][T_C]}{1 + k_{ST1}[T_C]}$$

with $k_S = f_S/b_S$, $k_{ST1} = f_{ST1}/b_S$. Assuming that $nk_S A \ll V$, $b_S/f_{ST2} \lesssim 1$, then $n[S]A$, $n[ST]A \ll [S_C]V$, and hence:

**Eq. 6**

$$[S_C] = \frac{N_S^{tot}}{V} = \frac{N_S^{tot}}{nV_0},$$

where we used the observation that the number of starch granules per unit volume inside the chloroplast is approximately constant, i.e., the number of granules $n$ is proportional to the chloroplast volume $V$ (Crumpton-Taylor et al., 2012). For the $T$ molecule, assuming the concentration in the compartment surrounding the granules is regulated by the circadian clock, we have (for much of the light/dark cycle, see Eq. 1):

$$[T_C] = \beta \Delta t .$$

The starch degradation rate for an individual granule is then:

$$\mu = m_S f_{D2} [S] A_d = m_S f_{D2} A_d k_S \frac{[S_C]}{1 + k_{ST1}[T_C]} .$$

According to Eq. 6, $[S_C]$ is proportional to the amount of starch in a granule, whereas from Eq. 7, $[T_C]$ is proportional to the expected time to next dawn. Sufficiently far from the time of expected dawn when $k_{ST1}[T_C] \gg 1$, we then have:

$$\mu = m_S f_{D2} A_d k_S \frac{\dfrac{N_S^{tot}}{nV_0}}{k_{ST1}\beta \Delta t} .$$

Hence, summing over all the granules, and using that $\Delta S^{tot} = \alpha N_S^{tot} = \dfrac{m_S f_{D2}}{f_{D1}} N_S^{tot}$ (see above in this section), the total starch degradation rate is:

$$\frac{d\Delta S^{tot}}{dt} = -\mu^{tot} = -\frac{f_{D1}A_d k_S}{k_{ST1}V_0 \beta} \frac{\Delta S^{tot}}{\Delta t} .$$

According to this equation, starch contents are completely depleted by the time of expected dawn, and for a degradation rate $\mu^{tot} = \dfrac{\Delta S^{tot}}{\Delta t}$ with normalization relation $\dfrac{f_{D1}A_d k_S}{k_{ST1}V_0 \beta} = 1$, starch contents decrease linearly with time.

## Model 2

In this model, division is now implemented by multiplying the $T$ concentration by the $S$ concentration, where the $T$ molecule concentration is proportional to $1/\Delta t$ for most of the light/dark cycle. Such a multiplication can be implemented by the following set of reactions:

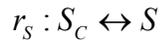

$r_S : S_C \leftrightarrow S$

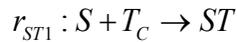

$r_{ST1} : S + T_C \rightarrow ST$

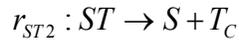

$r_{ST2} : ST \rightarrow S + T_C$

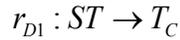

$r_{D1} : ST \rightarrow T_C$

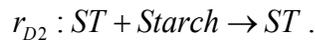

$r_{D2} : ST + Starch \rightarrow ST .$

Reversible reaction $r_S$ again describes the exchange of $S$ molecules between the surrounding compartment and the granule surface (forward reaction parameter $f_S$ and backward reaction parameter $b_S$). $S$ can recruit $T$ molecules from the surrounding compartment and form the complex $ST$ (reaction $r_{ST1}$, reaction parameter $f_{ST}$). As the $ST$ complex dissociates, $T$ is released back in the stroma, while $S$ remains attached to the granule (reaction $r_{ST2}$, reaction parameter $b_{ST}$). The $ST$ complex permits $S$ degradation, with $T$ released from the granule surface (reaction $r_{D1}$, reaction parameter $f_{D1}$), as well as starch degradation (reaction $r_{D2}$, reaction parameter $f_{D2}$). Similar to our assumptions for Model 1, we assume here that $b_{ST} \gg f_{D1}$, with a fixed degradation area $A_d$ for the starch as well as for the $S$ molecule as each granule shrinks. We also again assume that the dynamics of $S$ and $T$ can be taken to be in quasi-steady-state and that all granules in a given chloroplast have approximately equal areas and volumes.

The differential equations describing the dynamics of the total amount of starch in a chloroplast, $\Delta S^{tot}$, and of the total number of $S$ molecules, $N_S^{tot}$, are now:

$$\frac{dN_S^{tot}}{dt} = n\frac{dN_S}{dt} = -nf_{D1}[ST]A_d$$

$$\frac{d\Delta S^{tot}}{dt} = n\frac{d\Delta S}{dt} = -nm_S f_{D2}[ST]A_d = -n\mu = -\mu^{tot} \quad .$$

The above relations again ensure that if $\alpha = m_S \dfrac{f_{D2}}{f_{D1}}$, then $\Delta S^{tot} = \alpha N_S^{tot}$ at all times during degradation. The total number of $S$ molecules is $N_S^{tot} = [S_C]V + n[S]A + n[ST]A$, where $V$ is the compartmental volume surrounding the granules, and where $A$ is the surface area of a granule. Using the above quasi-steady state assumption and the law of mass action, we obtain:

$$[S] = k_S[S_C] \quad , \quad [ST] = k_{ST}k_S[S_C][T_C],$$

with $k_S = f_S/b_S$ and $k_{ST} = f_{ST}/b_{ST.}$ We assume that $nk_S A \ll V$, $k_{ST}[T_C] \lesssim 1$, giving $n[S]A$, $n[ST]A \ll [S_C]V$, and hence:

$$[S_C] = \frac{N_S^{tot}}{V} = \frac{N_S^{tot}}{nV_0},$$

where we again used the observation that the number of starch granules per unit volume inside the chloroplast is approximately constant, i.e., the number of granules $n$ is proportional to the volume $V$. For the $T$ molecules, we have (for much of the light/dark cycle, see Eq. 2):

$$[T_C] = \beta / \Delta t.$$

The starch degradation rate for an individual granule is therefore:

$$\mu = m_S f_{D2}[ST]A_d = m_S f_{D2} \; A_d k_{ST} k_S[S_C][T_C].$$

According to the above equation, $[T_C]$ is proportional to the reciprocal of the expected time to the next dawn. Thus, we have:

$$\mu = m_S f_{D2} \; A_d k_{ST} k_S \beta \frac{\dfrac{N_S^{tot}}{nV_0}}{\Delta t} \; .$$

Summing over all the granules, and using $\Delta S^{tot} = \alpha N_S^{tot} = \dfrac{m_S f_{D2}}{f_{D1}} N_S^{tot}$, the total starch degradation rate is

$$\frac{d\Delta S^{tot}}{dt} = -\mu^{tot} = -f_{D1} A_d \beta \frac{k_{ST} k_S}{V_0} \frac{\Delta S^{tot}}{\Delta t} .$$

According to this equation, starch contents are completely depleted by the time of expected dawn, and for a degradation rate $\mu^{tot} = \dfrac{\Delta S^{tot}}{\Delta t}$ with normalization relation $f_{D1} A_d \beta \dfrac{k_{ST} k_S}{V_0} = 1$, starch contents decrease linearly with time.

## **Calculation of the starch content during the degradation process**

In this section we describe how starch contents during the degradation process were calculated using the equations previously derived.

### **Model 1**

The starch content as function of time is obtained by solving the equation $\dfrac{d\Delta S^{tot}}{dt} = -\mu^{tot}$ . For Model 1, from the previous section, we have:

$$\mu^{tot} = m_S f_{D2} A_d k_S n \frac{[S_C]}{1 + k_{ST1}[T_C]} .$$

As we showed, $[S_C]$ can, at all times, be approximated by $[S_C] = \dfrac{N_S^{tot}}{n V_0} = \dfrac{\Delta S^{tot}}{\alpha n V_0} = \dfrac{f_{D1}}{m_S f_{D2} n V_0} \Delta S^{tot}$ ; therefore, we find:

$$\frac{d\Delta S^{tot}}{dt} = -f_{D1} A_d \frac{k_S}{V_0} \frac{\Delta S^{tot}}{1 + k_{ST1}[T_C]} .$$

If we put $\gamma = \dfrac{f_{D1} A_d k_s}{k_{ST1} V_0 \beta}$ , we then have

$$\frac{d\Delta S^{tot}}{dt} = -\gamma \frac{\beta k_{ST1} \Delta S^{tot}}{1 + k_{ST1}[T_C]} .$$

To convert this equation into one for starch content $\rho_S$ in $mg\ g^{-1}\ FW$, we use $\rho_S = (\Delta S^{tot}/n)\ v$, $(\Delta S^{tot}/n)$ being the amount of starch in a single granule in mg and $v$ the number of starch granules $g^{-1}\ FW$. This leads to

$$\frac{d\rho_S}{dt} = -\gamma \frac{\rho_S}{(\beta k_{ST1})^{-1} + ([T_C]/\beta)} .$$

Given Eq. 1 for $[T_C]$, the above differential equation can be exactly solved analytically. If we define the functions:

$$f^{(1)}(x,y) = \left( \frac{(\beta k_{ST1})^{-1} + \frac{t_{day} - t_r}{t_r} \, y}{(\beta k_{ST1})^{-1} + \frac{t_{day} - t_r}{t_r} \, x} \right)^{-\gamma \frac{t_r}{t_{day} - t_r}},$$

$$f^{(2)}(x,y) = \left( \frac{(\beta k_{ST1})^{-1} + t_{day} - y}{(\beta k_{ST1})^{-1} + t_{day} - x} \right)^{\gamma},$$

and $t^*$ is the time at which degradation starts, with $\rho_S(t^*) = \rho_0$ the corresponding starch content, then, for $0 \le t^* \le t_r$, the solution is:

$$\rho_S(t) = \begin{cases} \rho_0 f^{(1)}(t^*, t) & t^* \le t \le t_r \\ \\ \rho_0 f^{(1)}(t^*, t_r) f^{(2)}(t_r, t) & t_r \le t \le t_{day} \end{cases},$$

while, for $t_r \le t^* \le t_{day}$, the solution becomes:

$$\rho_S(t) = \begin{cases} \rho_0 f^{(2)}(t^*, t) & t^* \le t \le t_{day} \\ \\ \rho_0 f^{(2)}(t^*, t_{day}) f^{(1)}(0, t - t_{day}) & t_{day} \le t \le t_{day} + t_r \end{cases}.$$

According to these equations, for $(\beta k_{ST1})^{-1}$ sufficiently small, starch contents are almost completely depleted by the time of expected dawn, and for $\gamma=1$, starch contents decrease linearly with time for most of the dark period, except around the time of expected dawn, when $[T_C]$ is being reset (see Figure 2B).

**Model 2**

Similarly, for Model 2, we have:

$$\frac{d\rho_S}{dt} = -\gamma \; \rho_S \; \frac{[T_C]}{\beta},$$

where $\gamma = f_{D1} A_d \beta \dfrac{k_{ST} k_S}{V_0}$ and $[T_c]$ is defined by Eq. 2. Then, given the three functions:

$$f^{(1)}(x,y) = \exp\left\{-\frac{\gamma}{2}(y-x)\left(\frac{2}{\tau_2} - \frac{y+x}{\tau_1^2}\right)\right\},$$

$$f^{(2)}(x,y) = \left(\frac{t_{day}-y}{t_{day}-x}\right)^{\gamma},$$

$$f^{(3)}(x,y) = \exp\left\{-\frac{\gamma}{2}(y-x)\left(\frac{2}{\tau_3} - \frac{y+x}{\tau_1^2}\right)\right\},$$

with

$$\frac{1}{\tau_1^2} = \frac{t_{r2}-t_{r1}}{(t_{day}+t_{r1}-t_{r2})(t_{day}-t_{r1})(t_{day}-t_{r2})},$$

$$\frac{1}{\tau_2} = \frac{1}{(t_{day}+t_{r1}-t_{r2})}\left(\frac{t_{day}-t_{r2}}{t_{day}-t_{r1}} + \frac{t_{r1}}{t_{day}-t_{r2}}\right),$$

$$\frac{1}{\tau_3} = \frac{1}{(t_{day}+t_{r1}-t_{r2})}\left(\frac{t_{day}+t_{r1}}{t_{day}-t_{r2}} - \frac{t_{r2}}{t_{day}-t_{r1}}\right),$$

if the initial condition is $\rho_S(t^*) = \rho_0$, and $0 \le t^* \le t_{r1}$, the exact solution is:

$$\rho_S(t) = \begin{cases} \rho_0 f^{(1)}(t^*,t) & t^* \le t \le t_{r1} \\[2mm] \rho_0 f^{(1)}(t^*,t_{r1})f^{(2)}(t_{r1},t) & t_{r1} < t \le t_{r2} \\[2mm] \rho_0 f^{(1)}(t^*,t_{r1})f^{(2)}(t_{r1},t_{r2})f^{(3)}(t_{r2},t) & t_{r2} < t \le t_{day} \end{cases}.$$

For $t_{r1} \le t^* \le t_{r2}$, we have:

$$\rho_S(t) = \begin{cases} \rho_0 f^{(2)}(t^*,t) & t^* \le t \le t_{r2} \\[2mm] \rho_0 f^{(2)}(t^*,t_{r2})f^{(3)}(t_{r2},t) & t_{r2} \le t \le t_{day} \\[2mm] \rho_0 f^{(2)}(t^*,t_{r2})f^{(3)}(t_{r2},t_{day})f^{(1)}(0,t-t_{day}) & t_{day} \le t \le t_{r1}+t_{day} \end{cases},$$

and for $t_{r2} \le t^* \le t_{day}$:

$$\rho_S(t) = \begin{cases} \rho_0 f^{(3)}(t^*, t) & t^* \leq t \leq t_{day} \\[2mm] \rho_0 f^{(3)}(t^*, t_{day}) f^{(1)}(0, t - t_{day}) & t_{day} \leq t \leq t_{r1} + t_{day} \\[2mm] \rho_0 f^{(3)}(t^*, t_{day}) f^{(1)}(0, t_{r1}) f^{(2)}(t_{r1}, t - t_{day}) & t_{r1} + t_{day} \leq t \leq t_{r2} + t_{day} \end{cases} .$$

From the previous equations, it is easily seen that for $t_{r2}$ close to $t_{day}$ starch contents are almost completely depleted by the time of expected dawn, and for $\gamma=1$, starch contents decrease linearly with time for $t_{r1} \leq t \leq t_{r2}$, although this linearity no longer holds around the time of expected dawn, when $[T_C]$ is being reset (see Figure 2C).

**Parameters**

For Model 1, the values of the following parameters are required: $\gamma$, $(\beta k_{STI})^{-1}$, $\rho_0$, $t_r$; and for Model 2: $\gamma$, $\rho_0$, $t_{r1}$ and $t_{r2}$.

In order to fit the data from the night-time light period experiments, we also considered the possibility that the $T$ dynamics given by Eqs. 1 and 2 (see also Figures 2B,C) can be phase shifted by a time $t_0$. The addition of $t_0$ as an extra fitting parameter for the night-time light period experiments was justified by our data on transcripts of the clock gene *LHY* (Figure 3 – Figure Supplement 1), showing that the night-time light period may induce a phase shift in the expression of some genes.

The phase shift of the $T$ dynamics determines a shift in the time of expected dawn, that, in Model 1, is defined as the time at which $[T_C]$ falls to zero (see Eq. 1 and Figure 2B) and which in Model 2 is the time at which $[T_C]$ would diverge without a reset (see Eq. 2 and Figure 2C). Therefore, if the $T$ dynamics are phase shifted by $t_0$, in our models the starch degradation rates are adjusted in such a way so as to deplete the starch reserves at $(24+t_0)$ h after the previous dawn, instead of the normal 24 h. We found this phenotype in *cca1/lhy* plants, which run out of starch earlier than 24 h after the previous dawn. Hence, to reproduce the phenotype of *cca1/lhy* in our models, $t_0$ was also used as a fitting parameter. A more extensive discussion about *cca1/lhy* can be found in the "Mutant phenotypes" section.

A full list of the system parameters is shown in Tables 1 and 2.

**Data Fitting**

The best fit to a given dataset was found by minimizing the function:

$$L(\vec{\theta}) = \sum_{i=1}^{N} \left( \frac{\rho_S^{theory}(t_i, \vec{\theta}) - \rho_S^{exp}(t_i)}{\rho_S^{exp}(t_i)} \right)^2 ,$$

where $\rho_S^{exp}(t_i)$ are the $N$ mean values of starch contents measured at times $t_i$. $\rho_S^{theory}(t_i, \vec{\theta})$ is the starch value predicted by the model at time $t_i$ with parameter values $\vec{\theta}$. The set of parameters $\vec{\theta}$ that minimizes $L$ corresponds to the maximum-likelihood estimates of the parameters under the assumption that the experimental measurements are normally distributed around the theoretical values with a constant relative error.

Note that, in the fits of the data from the *cca1/lhy* mutant plants for the early and the normal night, we did not consider the two data points closest in time to 24 h from the datasets, as they are characterized by very low values of starch content and, therefore, are likely to be affected by higher relative errors compared to the other data points. For the same reason, for the linear fits shown in Figure 4 – Figure Supplement 1A we did not consider the data point at t=24 h in the normal night datasets and the data points at t=22 h and t=24 h in the early night datasets.

A simulated annealing algorithm was used for the minimization of $L(\vec{\theta})$. The parameter values of the best fits of the models are given in Tables 3-6, along with the ranges in which the parameters were allowed to vary.

### Recalculation of the starch degradation rate after the night-time light period

By combining our results from the night-time light period experiments shown in Figure 3A-C, we can show that the evidence in favor of a re-calculation of the starch degradation rate after the night-time light period is statistically significant. In the experiments shown in Figure 3A-C, plants were subjected to 5 h of light in the middle of the night (between 14 h and 19 h after the previous dawn). The starch content at each time point was measured by averaging the starch content of n=10-12 individual rosettes and the standard error of the mean was also calculated. The starch degradation rate before the night-time light period in the *i-th* experiment was measured as:

$$\mu_B^i = \frac{\rho_S^{\ i}(12h) - \rho_S^{\ i}(14h)}{14h - 12h},$$

i.e., as the difference between the starch content at 12 h and 14 h divided by the time interval. Similarly, the starch degradation rate after the night-time light period for the *i-th* experiment was:

$$\mu_A^i = \frac{\rho_S^{\ i}(19h) - \rho_S^{\ i}(21h)}{21h - 19h},$$

These rates were averaged over the three experiments, and the difference between the two averages was calculated:

$$\mu_A - \mu_B = \frac{1}{N}\sum_{i=1}^{N}\left(\mu_A^i - \mu_B^i\right),$$

where *N*=3 was the number of experiments. The error on this quantity was estimated by propagating standard errors of the mean. We found $\mu_A - \mu_B$ = 0.43±0.14 *(mg g⁻¹ FW)/h* which is significantly greater than 0 (one-tailed Welch's t-test, p-value $=1.6\times10^{-3}$).

We also calculated the average difference between $\mu_A^i$ and

$$\mu_{normal}^i = \frac{\rho_S^{\ i}(12h)}{12h},$$

where $\mu_{normal}^i$ is the rate that would have been expected over a normal 12-h night. We found that this average is

$$\mu_A - \mu_{normal} = \frac{1}{N}\sum_{i=1}^{N}\left(\mu_A^i - \mu_{normal}^i\right) = 0.48 \pm 0.11 \left(mg\,g^{-1}FW\right)/h\,,$$

where the error was again estimated by propagating the standard errors of the mean.
This value is again significantly greater than 0 (one-tailed Welch's t-test, p-value = $3.8\times10^{-5}$).

These results provide strong evidence against the hypothesis that a fixed degradation rate is set only once at the first onset of darkness, and is instead compatible with our prediction that the rate is continuously re-computed throughout the night.

## **Mutant phenotypes**

In the following, we show how the models can explain the mutant phenotypes discussed in the main text (see Figure 1B, Figure 4 and Figure 4 – Figure Supplement 1).

### *cca1/lhy*

LHY and CCA1 are central components of the clock. The *cca1/lhy* mutant has a free-running period of significantly less than 24 h under continuous light.

This mutant is characterized by too high a rate of starch degradation (see Figure 1B). Indeed, the starch reserve is exhausted around 21-22 h after the previous dawn, instead of 24 h as in the wild type. Interestingly, if these mutant plants are given an early night, the starch degradation rate is adjusted such that all the starch is again degraded around 21-22 h after previous dawn (see Figure 1B).

The models we discussed previously can straightforwardly explain such a phenotype by assuming that in this mutant, the time of expected dawn is shifted to a time t < 24 h after the previous dawn. There are different perturbations of the $T$ dynamics that can produce this effect in our models, and still reproduce the data from *cca1/lhy* equally well. For instance, for *cca1/lhy* in Model 1, $[T_C]$ after being reset, could decrease more steeply than in the wild-type, and drop to zero at a time t < 24 h after the previous dawn. Then $[T_C]$ could remain at low levels, before rising again around 24 h. Another perturbation appropriate for both Models 1 and 2, would be to phase shift the $T$ dynamics given by Eqs. 1 and 2 respectively by a time $t_0 < 0$ h, in such a way that the time of expected dawn becomes $(24+t_0)$ h< 24 h, as discussed above in the "Parameters" section. For the sake of simplicity, we fitted the *cca1/lhy* data by assuming that the latter perturbation takes place, and accordingly we used $t_0$ as a fitting parameter.

### *lsf1, sex4, bam3, bam4, isa3*

*lsf1* and *sex4* mutant plants were kept in the dark for 132 h after the end of the previous light period, then exposed to normal 12 h light,12 h dark cycles and the starch content measured. As Figure 4A shows, during the days following the prolonged dark period the mutant plants failed to degrade their entire starch reserve at night. Instead, the total starch content degraded by the end of each night, expressed as a percentage of the starch content at the end of the respective preceding light period, was approximately constant and much lower than that of wild-type plants (Figure 4B).

Interestingly, the same mutants could also adjust their starch degradation rate in response to an unexpected early night. In order to quantitatively verify this observation, we performed linear fits of the data from the unexpected early and normal nights:

$$\rho_S^{early}(t) = -\mu^{early} \cdot (t - t^{early}) + \rho_{S,0}^{early},$$

$$\rho_S^{normal}(t) = -\mu^{normal} \cdot (t - t^{normal}) + \rho_{S,0}^{normal},$$

where ($\mu^{early}$, $\rho_{S,0}^{early}$) and ($\mu^{normal}$, $\rho_{S,0}^{normal}$) are the fitting parameters, with $t^{early}$=8 h, $t^{normal}$=12 h the times of onset of darkness respectively for the unexpected early and normal night. From these fits, we calculated the ratio $R$ between the starch degradation rates in the normal and the unexpected early night, normalized by the respective starch content at the time of onset of darkness:

$$R = \frac{\mu^{normal} / \rho_{S,0}^{normal}}{\mu^{early} / \rho_{S,0}^{early}}.$$

If the degradation rate is adjusted as happens in wild-type plants, during the unexpected early night relative to a normal night, the same fraction $f$ of the initial starch content should be degraded by the time of expected dawn. Therefore, $\mu^{normal} \approx f \dfrac{\rho_{S,0}^{normal}}{t_{day} - t^{normal}}$ and

$\mu^{early} \approx f \dfrac{\rho_{S,0}^{early}}{t_{day} - t^{early}}$; hence, $R \approx \dfrac{t_{day} - t^{early}}{t_{day} - t^{normal}} = \dfrac{16}{12} \approx 1.33$. As Figure 4F shows, the values of $R$ found for *lsf1* and *sex4* are compatible with 1.33 (dashed line), therefore supporting the hypothesis that these mutants are able to appropriately adjust their starch degradation rate in response to an unexpected early night. We also found the same type of rate adjustment in *bam3, bam4* and *isa3* mutants (see Figure 4F and Figure 4 – Figure Supplement 1), which also failed to exhaust their starch reserves by the end of the night.

The models make a precise prediction about how such a phenotype can be produced. We will focus on Model 1, but analogous arguments also hold true for Model 2. We showed that for Model 1, in order to obtain a linearly decreasing starch content with time, and with the appropriate degradation rate to ensure complete starch depletion at the time of expected dawn, two normalization conditions must hold:

$$\frac{f_{D1} A_d k_S}{k_{ST1} V_0 \beta} = \gamma = 1, \quad \frac{m_S f_{D2}}{\alpha f_{D1}} = \chi = 1.$$

The first condition ensures that the number of $S$ molecules, and therefore the starch content, decreases linearly with time during the night. The second condition is needed to keep the number of $S$ molecules and the amount of starch strictly proportional, so that both these quantities are fully degraded by the end of the night. We now assume that the second of the two conditions is not valid, and, in particular, that $\chi < 1$. In this case, the starch degradation proceeds at a slower rate, breaking the proportionality between the number of S molecules and the amount of starch. By using Eqs. 4 and 8, and the initial condition at the start of the dark period, $\Delta S^{tot} = \alpha N_S^{tot}$, we find that a fraction of starch approximately equal to $\chi$ only is degraded by the time of expected dawn. Such a phenotype is clearly compatible with the disruptions caused by the above mutations. Indeed by assuming that the second condition above is not valid, our models can be well fit to the full night-time starch profiles in all the above mutants (data not shown).

One way to perturb only the second of the two normalization conditions shown above (and not the first), would be to alter $f_{D2}$, which is the starch degradation reaction parameter. This reasoning fits well with the known functions of the above genes, all of which play roles, directly or indirectly, in starch degradation.

### *pwd*

The *pwd* mutant lacks the glucan water dikinase responsible for the addition of the phosphate groups to the 3-position of glucose moieties in starch. This mutation generated an approximately linear decrease of starch content with time during the night, but with a rate which was too low to ensure the complete utilization of the starch reserve by the time of expected dawn. Yet, as opposed to the previously discussed mutations, this mutant did not have the ability to adjust the starch degradation rate in response to an unexpected early night (see Figure 4E,F). Indeed, for this mutant we found that the ratio between the degradation rates (normalized by the respective end-of-light period starch content) during the normal and the early night was $R=1.10\pm0.10$, which is significantly less than the wild-type value ($R\approx1.33$, one-tailed Z-test, p-value = 0.01).

This finding indicates that PWD is potentially a key node where information about starch content and expected time to dawn from the circadian clock are integrated to control starch degradation dynamics. This makes PWD an obvious target for future experiments that aim to elucidate the identities of the $S$ and $T$ molecules.

## Other examples of models implementing arithmetic division

The models detailed above use two distinctly different methods of implementing the division operation. In one case, exemplified by Model 1, the $S$ and $T$ molecule concentrations are divided, with the $T$ molecule concentration tracking the time to expected dawn. In Model 2, the $S$ and $T$ concentrations are multiplied, with the $T$ molecule concentration tracking the reciprocal of the time to expected dawn. The precise implementation of these two methods of performing arithmetic division could, however, vary, with slightly different reaction schemes but the same underlying principles. To illustrate this point, we briefly outline other models related to Model 1 above, where the $T_C$ molecule again has the behavior given in Eq. 1. We first consider:

$$r_S : S_C \leftrightarrow S$$

$$r_T : T_C \leftrightarrow T$$

$$r_{T_2} : T + T \leftrightarrow T_2$$

$$r_{ST} : S + T \leftrightarrow ST$$

$$r_{ST_2} : S + T_2 \rightarrow ST_2$$

$$r_{S_CT_2} : ST_2 \rightarrow S_C + T_2$$

$$r_{D1} : ST \rightarrow T$$

$$r_{D2} : ST + Starch \rightarrow ST.$$

Here both $S$ and $T$ molecules can directly and reversibly associate to the granule surface (reactions $r_S$ and $r_T$). Once bound to the surface, $S$ and $T$ can form a complex $ST$ (reaction $r_{ST}$) that permits $S$ molecule and starch degradation (through reactions $r_{D1}$ and $r_{D2}$ respectively). The $T$ molecule can also hinder starch degradation by forming dimers (reaction $r_{T_2}$) that are able to associate with $S$ molecules (reaction $r_{ST_2}$) and induce them to detach from the granule surface (reaction $r_{S_cT_2}$). By a similar analysis to that carried out for Model 1 and Model 2, and for similar reasons, it can be shown that this model can also implement arithmetic division between the $S$ and $T$ molecule concentrations.

More complex models can also easily accommodate additional molecules, which, for instance, could recruit $S$ and $T$ molecules to the granule surface and be part of the starch degradation apparatus. For example, in the set of reactions:

$r_M : M_C \leftrightarrow M$

$r_S : S_C + M \rightarrow SM$

$r_T : T_C + M \rightarrow TM$

$r_{SM} : SM \rightarrow S_C + M_C$

$r_{TM} : TM \rightarrow T_C + M_C$

$r_{D1} : SM \rightarrow M_C$

$r_{D2} : SM + Starch \rightarrow SM$,

$M$ molecules which reversibly bind to the granule surface (reaction $r_M$), recruit the $S$ and $T$ molecules from the surrounding compartment (reactions $r_S$ and $r_T$), to form the $SM$ and $TM$ complexes, which can then dissociate from the surface through the reactions $r_{SM}$ and $r_{TM}$. The $SM$ complex can permit $S$ molecule and starch degradation (reactions $r_{D1}$ and $r_{D2}$). Since the $T$ molecule hinders starch degradation by binding to $M$ molecules and preventing them from binding to $S$, it can be shown that this model can also implement arithmetic division between the $S$ and $T$ molecule concentrations.

## Validation of the use of a night-time light period

Exposure of Arabidopsis plants to light during the normal night could potentially lead to the early, light-induced expression of clock genes, and under some circumstances might re-entrain the clock. Such re-entrainment might affect the level of the $T$ molecule. To discover whether our experimental conditions gave rise to such problems, we investigated the behavior of transcript levels of *LHY*, a central clock gene, by qPCR analysis. For the experiment shown in Figure 3B, *LHY* transcript levels were slightly elevated during the night-time light period, then peaked about 2h later than in plants that were not exposed to the light period (Figure 3 – Figure Supplement 1). This result indicates that the night-time light period has only minor effects on the clock, which is not re-entrained: if this were the case the peak of *LHY* expression observed with the night-time light period would not be expected to occur at around 24h after the previous dawn.

## Acknowledgements

We thank Mark Stitt and members of the Howard group for discussions.

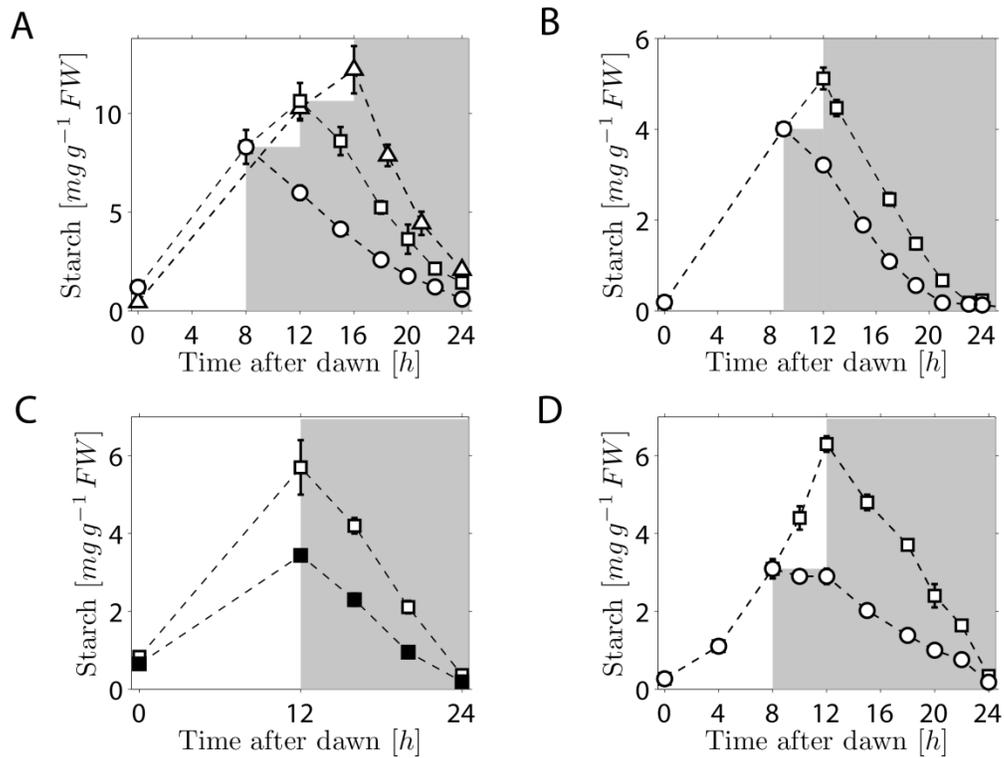

**Figure 1**.

**Starch content levels from experiments with unexpected variation in either starch content at the onset of darkness or the time of onset of darkness.** (A) Starch turnover in Arabidopsis grown in 12 h light, 12 h dark, then subject to unexpected early (8 h, n=6 individual rosettes, circles) normal (12 h, n=6, squares) or unexpected late (16 h, n=5, triangles) onset of darkness. (B) Starch turnover in Arabidopsis *cca1/lhy* mutant grown in 12 h light, 12 h dark, then subject to unexpected early (9 h, circles), or normal (12 h, squares) onset of darkness (n=6-10). (C) Starch turnover in Arabidopsis exposed to different daytime light levels: 90 µmol quanta m$^{-2}$ s$^{-1}$ (open squares) or 50 µmol quanta m$^{-2}$ s$^{-1}$ (filled squares) (both n=5, previously all plants grown in 12 h light, 12 h dark with 90 µmol quanta m$^{-2}$ s$^{-1}$). (D) Starch turnover in *Brachypodium* grown in 12 h light, 12 h dark, then subject to unexpected early (8 h, circles) or normal (12 h, squares) onset of darkness (both n=6). Error bars are standard error of the mean throughout.

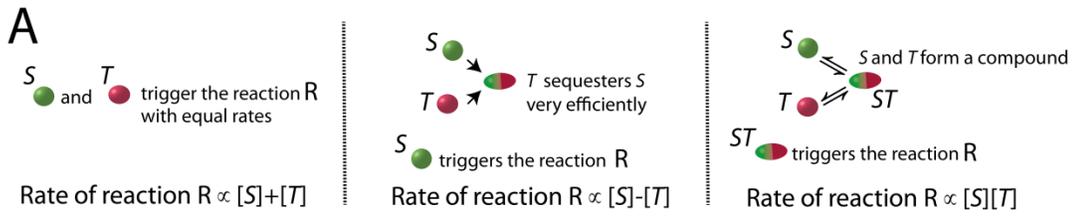

**A**

$S$ and $T$ trigger the reaction R with equal rates

Rate of reaction R $\propto [S]+[T]$

$T$ sequesters $S$ very efficiently

$S$ triggers the reaction R

Rate of reaction R $\propto [S]-[T]$

$S$ and $T$ form a compound

$ST$ triggers the reaction R

Rate of reaction R $\propto [S][T]$

**B**
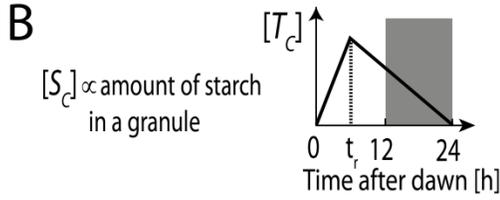

$[S_C] \propto$ amount of starch in a granule

**C**
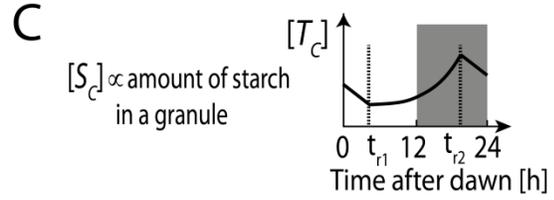

$[S_C] \propto$ amount of starch in a granule

**D**
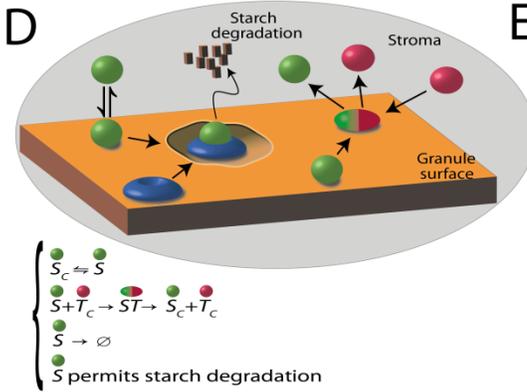

$\begin{cases} S_C \approx S \\ S + T_C \to ST \to S_C + T_C \\ S \to \emptyset \end{cases}$

$S$ permits starch degradation

**E**
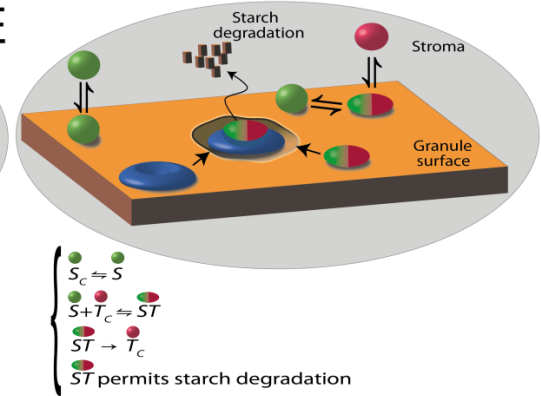

$\begin{cases} S_C \approx S \\ S + T_C \approx ST \\ ST \to T_C \end{cases}$

$ST$ permits starch degradation

**F**
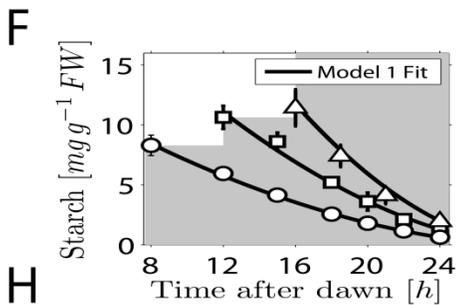

**G**
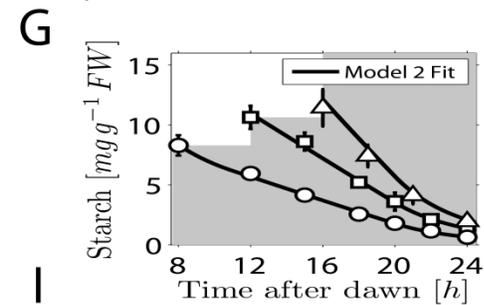

**H**
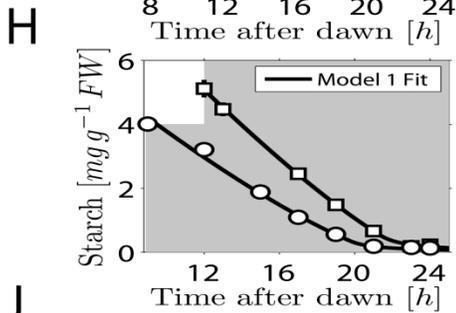

**I**
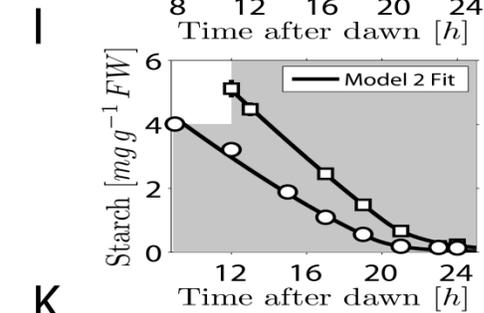

**J**
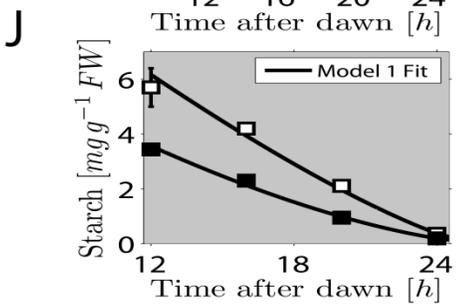

**K**
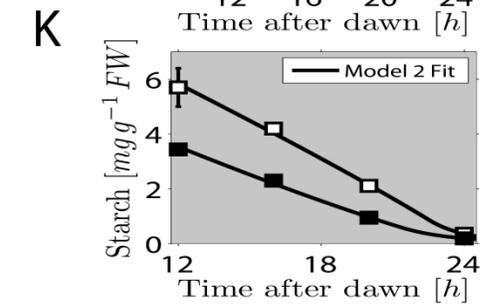

**Figure 2.**

**Chemical kinetic models capable of implementing analog arithmetic operations.** (A) Pictorial summaries of schemes for analog implementation of addition, subtraction and multiplication between the concentrations of two molecules $S$ and $T$. Square brackets indicate concentrations. (B,C) Schematic behavior of the stromal concentrations of $S$ and $T$ molecules ([$S_C$] and [$T_C$] respectively), in (B) first and (C) second arithmetic division models. In the first model, the $T$ molecule tracks the time to expected dawn after a reset-time $t_r$. In the second model the $T$ molecule concentration increases with time proportionally to 1/(expected time to dawn) between $t_{r1}$ and $t_{r2}$. (D,E) Pictorial summaries of (D) first and (E) second analog arithmetic division models (not all reactions shown in pictures, for full details see Materials and Methods). In the reaction schemes, molecules not attached to the starch granule surface have a "$C$" subscript. The blue disk represents components of the starch degradation apparatus potentially activated by the $S$ molecule in the first model, and by the $ST$ complex in the second model. Best fits (full lines) of first (F,H,J) and second (G,I,K) arithmetic division models to Arabidopsis data from Figure 1A,B,C.

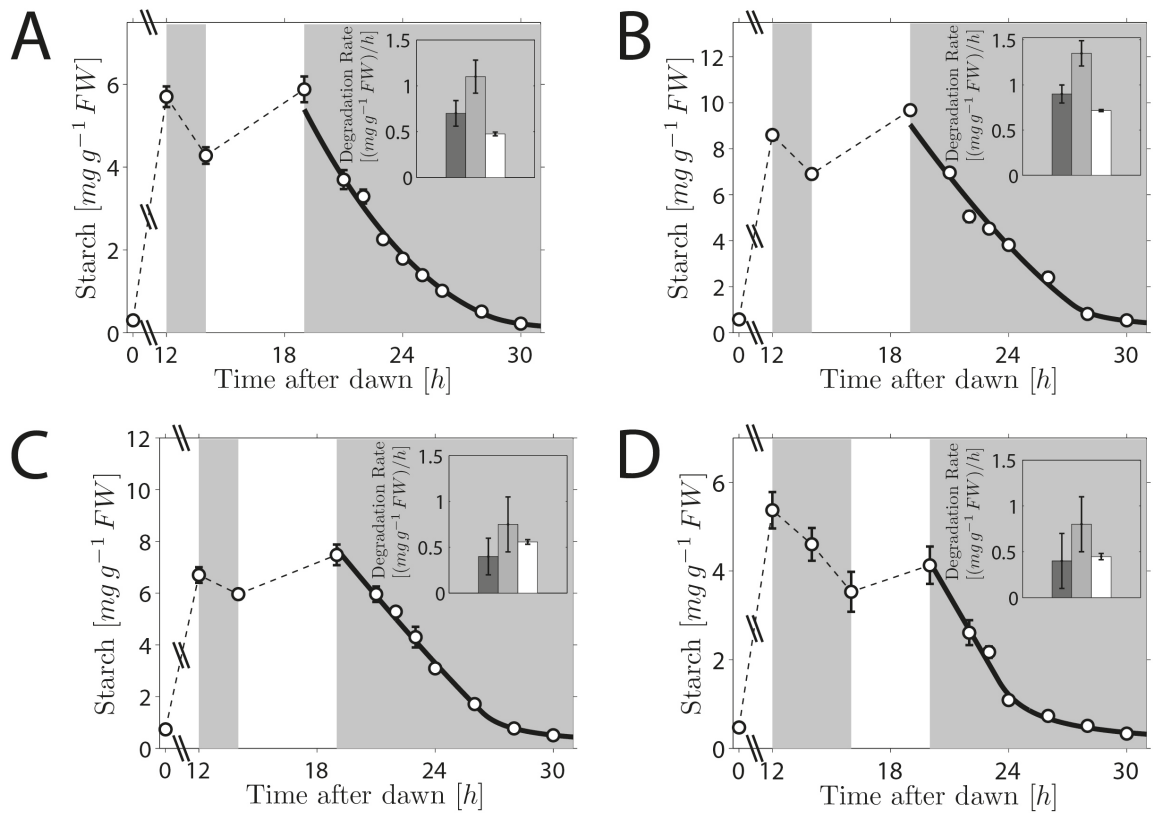

**Figure 3.**

**Starch content levels from experiments incorporating night-time light period.** Arabidopsis plants grown in 12 h light, 12 h dark were subjected to onset of darkness at 12 h, followed by an unexpected period of light, followed by extended darkness. (A,B,C) Three data sets (n=12 individual rosettes, except n=10 for (C)), in which the unexpected period of light was between 14 h and 19 h after dawn. (D) In the fourth dataset (n=12) the period of light was between 16 h and 20 h after dawn. Full lines are best fits to the first division model. The second model produces very similar fits (see Figure 3 – Figure Supplement 2).The insets show the respective starch degradation rates computed from the 12 h and 14 h experimental time points (dark grey bars) compared to those computed from the 19 h and 21 h experimental time points in panels (A,B,C) or the 20 h and 22 h time points in panel (D) (light grey bars). The white bars are the expected starch degradation rates in a normal 12 h night, i.e. rates that would have ensured the complete depletion of the starch content measured at 12 h at the time of expected dawn (24 h). Error bars are standard error of the mean throughout.

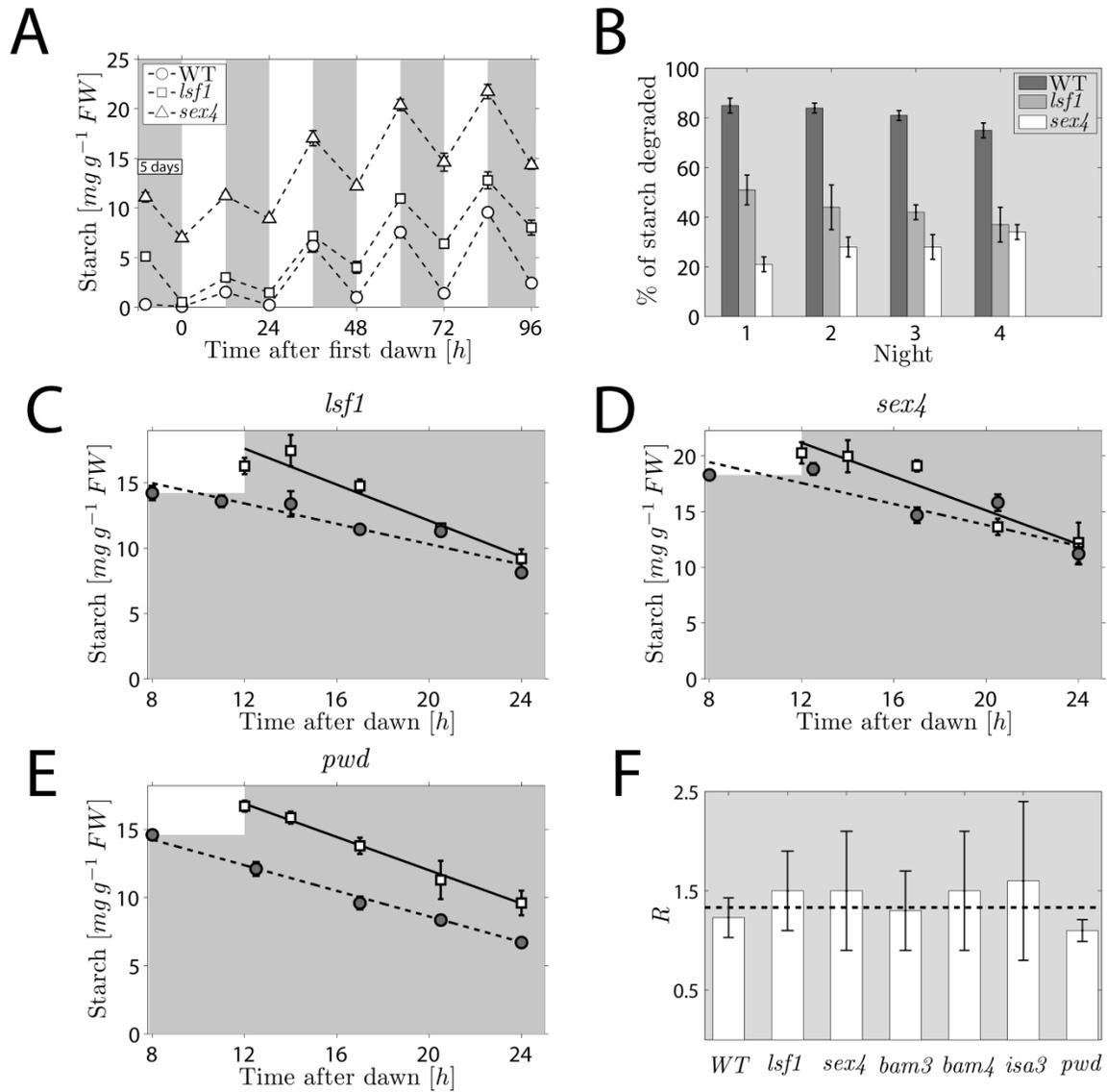

**Figure 4.**

**Starch content levels in mutant Arabidopsis plants defective in components of the starch degradation apparatus.** (A) Starch content in wild-type (WT) plants and *lsf1* and *sex4* mutant plants during four days of 12 h light, 12 h dark following five days of continuous darkness, where plants were transferred back into the light (at time 0 h on the x-axis) 132 h after the end of the previous light period (n=6 individual rosettes). Data for wild-type and *lsf1* plants are from (Comparot-Moss et al., 2010). (B) The percentage of starch degraded during each of the 4 nights in (A). (C,D,E) Starch content in *lsf1, sex4* and *pwd* mutant plants grown in 12 h light, 12 h dark cycles then subject to unexpected early (8 h, circles) or normal (12 h, squares) onset of darkness (n=5). The continuous and dashed lines are linear fits to the normal and early night datasets respectively. (F) For each of the labeled genotypes, $R$ is the ratio between the starch degradation rates (each normalized by their respective end-of-light period starch content and as determined from the linear fits) during the normal and early nights. The dashed line shows the expected value of $R$ for wild-type (WT) plants, i.e. ratio of rates that would ensure the complete

depletion of the starch content in all cases at the time of expected dawn (24 h). See Materials and Methods for details about the linear fitting and the calculation of $R$. Error bars are standard error of the mean throughout. Figure 4 – Figure Supplement 1 shows the datasets used to calculate $R$ for WT, *bam3, bam4* and *isa3.*

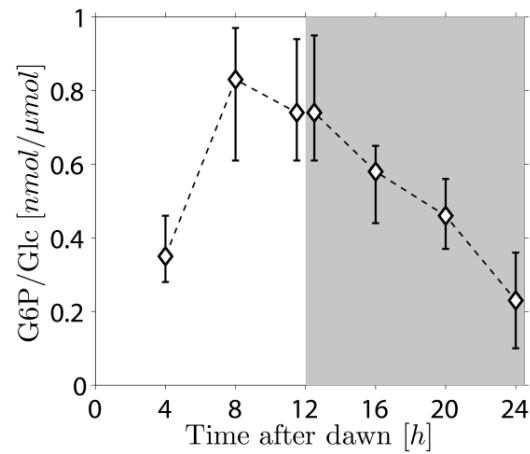

**Figure 5.**

**Daily change in starch phosphate content (measured as glucose 6-phosphate, G6P) in Arabidopsis leaves.** Results are normalized by total amount of glucose (Glc) in starch at each time point. Starch was extracted from rosettes of 26-day-old plants. n=3 pools of 10 rosettes except at 24 h time point, with n=2 pools of 15 rosettes. Error bars represent the range (i.e. error bar edges correspond to highest and lowest values measured).

**TABLES**

**Table 1.**

Full list of the system parameters for Model 1.

| Symbol | Definition |
|--------|------------|
| $\rho_0$ | Starch content at the beginning of the dark period. |
| $\gamma$ | Normalization variable |
| $(\beta k_{ST1})^{-1}$ | $k_{ST1}$ is the ratio of the reaction parameter associated with reaction $r_{ST1}$ with the backward rate of reaction $r_S$. $\beta$ is the proportionality constant between $[T_C]$ and $\Delta t$. |
| $t_r$ | Time at which $[T_C]$ levels finish being reset at the beginning of the day (see Figure 2B). |
| $t_0$ | Phase shifting parameter of the $[T_C]$ dynamics given by Eq. 1. The next dawn is expected to come $(24+t_0)$ h after the previous one. |

**Table 2.**

Full list of the system parameters for Model 2.

| Symbol | Definition |
|--------|------------|
| $\rho_0$ | Starch content at the beginning of the dark period. |
| $\gamma$ | Normalization variable |
| $t_{r1}$ | Time at which $[T_C]$ levels finish being reset at the beginning of the day (see Figure 2C). |
| $t_{r2}$ | Time at which $[T_C]$ levels start being reset at the end of the day (see Figure 2C). |
| $t_0$ | Phase shifting parameter of the $[T_C]$ dynamics given by Eq. 2. The next dawn is expected to come $(24+t_0)$ h after the previous one. |

**Table 3.**

The values of the parameters of the Model 1 best fits shown in Figure 2. Below each parameter the range used in the best fit search is indicated. WT indicates wild-type plants.

| | Model 1 | | | | | | |
|---|---|---|---|---|---|---|---|
| | **WT Early Night (Panel F)** | **WT Normal Night (Panel F)** | **WT Late Night (Panel F)** | *cca1/lhy* **Early Night (Panel H)** | *cca1/lhy* **Normal Night (Panel H)** | **WT Low Light Level (Panel J)** | **WT Normal Light Level (Panel J)** |
| $\rho_0$ *[mg g$^{-1}$ FW]* Within 10% of the measured value | 8.5 | 11.0 | 11.7 | 4.2 | 5.1 | 3.6 | 6.1 |
| $\gamma$ [0.7 - 3.0] | 1.8 | 1.8 | 1.9 | 1.2 | 1.2 | 1.5 | 1.3 |
| $(\beta k_{ST1})^{-1}$ [1.5 − 5.0] h | 5.0 | 5.0 | 5.0 | 1.6 | 1.7 | 2.1 | 1.5 |
| $t_r$ [9.0 − 12.0] h | 9.0 | Any value in the specified range | Any value in the specified range | 11.7 | 11.7 | Any value in the specified range | Any value in the specified range |
| $t_0$ [(-5.0) − 5.0] h for the *cca1/lhy* data, $t_0$=0 for WT | 0.0 | 0.0 | 0.0 | -4.2 | -2.5 | 0.0 | 0.0 |

**Table 4.**

The values of the parameters of the Model 2 best fits shown in Figure 2. Below each parameter the range used in the best fit search is indicated. WT indicates wild-type plants.

| | Model 2 | | | | | | |
|---|---|---|---|---|---|---|---|
| | WT Early Night (Panel G) | WT Normal Night (Panel G) | WT Late Night (Panel G) | *cca1/lhy* Early Night (Panel I) | *cca1/lhy* Normal Night (Panel I) | WT Low Light Level (Panel K) | WT Normal Light Level (Panel K) |
| $\rho_0$ *[mg g$^{-1}$ FW]* Within 10% of the measured value | 8.3 | 11.0 | 12.1 | 4.2 | 5.1 | 3.5 | 5.8 |
| $\gamma$ [0.7 -3.0] | 1.1 | 1.0 | 1.1 | 1.3 | 1.2 | 1.2 | 0.9 |
| $t_{r1}$ [9.0 – 12.0] h | 11.0 | 12.0 | 12.0 | 11.0 | 11.4 | 10.5 | 10.2 |
| $t_{r2}$ [20.0 – 23.0] h | 20.1 | 20.0 | 20.0 | 21.8 | 21.4 | 21.4 | 22.5 |
| $t_0$ [(-5.0) – 5.0] h for the *cca1/lhy* data, $t_0$=0 for WT | 0.0 | 0.0 | 0.0 | -2.4 | -1.3 | 0.0 | 0.0 |

**Table 5.**

The values of the parameters of the Model 1 best fits shown in Figure 3. Below each parameter the range used in the best fit search is indicated.

| | Model 1 | | | |
|---|---|---|---|---|
| | **Panel A** | **Panel B** | **Panel C** | **Panel D** |
| $\rho_0$ *[mg g$^{-1}$ FW]* Within 10% of the measured value | 5.4 | 9.0 | 7.7 | 4.2 |
| $\gamma$ [0.7 − 3.0] | 2.4 | 1.4 | 1.1 | 1.0 |
| $(\beta k_{ST1})^{-1}$ [1.5 − 5.0] h | 5.0 | 2.6 | 2.1 | 2.0 |
| $t_r$ [9.0 − 12.0] h | 9.0 | 9.0 | 10.3 | 12.0 |
| $t_0$ [(-5.0) − 5.0] h | 4.3 | 3.3 | 2.4 | -0.5 |

**Table 6.**

The values of the parameters of the Model 2 best fits shown in Figure 3 – Figure Supplement 2. Below each parameter the range used in the best fit search is indicated.

| | Model 2 | | | |
|---|---|---|---|---|
| | **Panel A** | **Panel B** | **Panel C** | **Panel D** |
| $\rho_0$  *[mg g$^{-1}$ FW]* <br> Within 10% of the measured value | 5.5 | 9.1 | 7.7 | 4.2 |
| $\gamma$ <br> [0.7 – 3.0] | 1.5 | 1.3 | 0.7 | 0.7 |
| $t_{r1}$ <br> [9.0 – 12.0] h | 12.0 | 9.0 | 9.0 | 9.0 |
| $t_{r2}$ <br> [20.0 – 23.0] h | 20.0 | 20.9 | 22.1 | 22.1 |
| $t_0$ <br> [(-5.0) – 5.0] h | 5.0 | 5.0 | 2.4 | 0.0 |

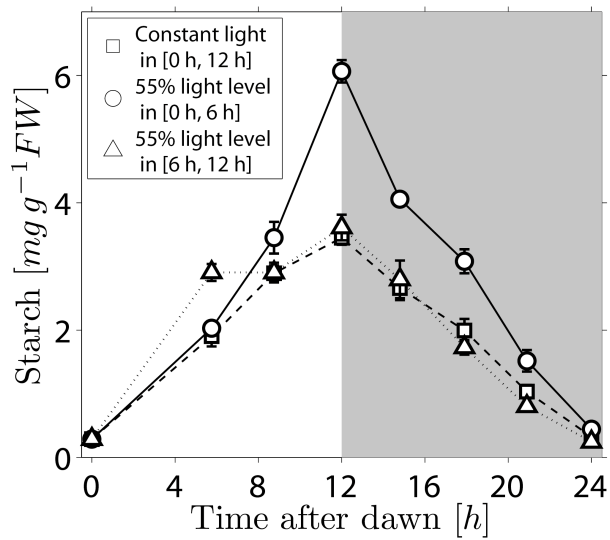

**Figure 1 – Figure Supplement 1.**

**Starch content levels in Arabidopsis plants exposed to different regimes of varying light level over a single light period.** Three sets of plants (each n=5 individual rosettes) were grown in 12 h light, 12 h dark and were then subject to different light regimes during a single day. One set (squares) was exposed to normal light levels (180 µmol quanta m$^{-2}$ s$^{-1}$), the other two were shaded to about 55% of normal light level (100 µmol quanta m$^{-2}$ s$^{-1}$) for either the first 6 h (circles) or the second 6 h (triangles) of the 12 h light period, with the normal light level for the other 6 h period. Error bars are standard error of the mean.

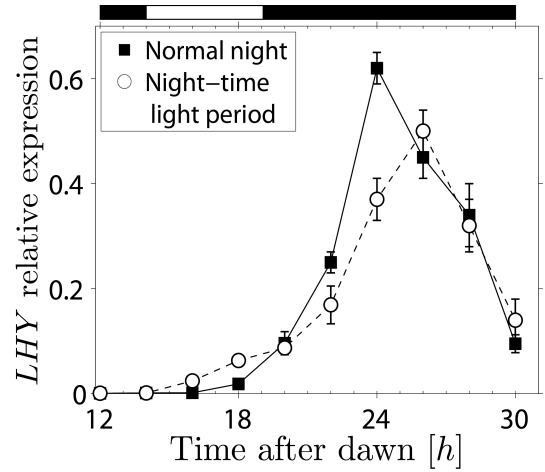

**Figure 3 – Figure Supplement 1.**

**Transcript levels of *LHY* from experiment incorporating night-time light period.** *LHY* transcript levels (relative to *ACT2*) measured in Arabidopsis plants kept in continuous darkness after a normal night (squares), or subjected to a 5 h night-time light period between 14 h and 19 h after dawn, and then kept in continuous darkness (circles), as in Figure 3A-C. Data for the night-time light period are from the same plants as in Figure 3B. n=5 individual rosettes, error bars are standard error of the mean. The night-time light period is shown on top of graph.

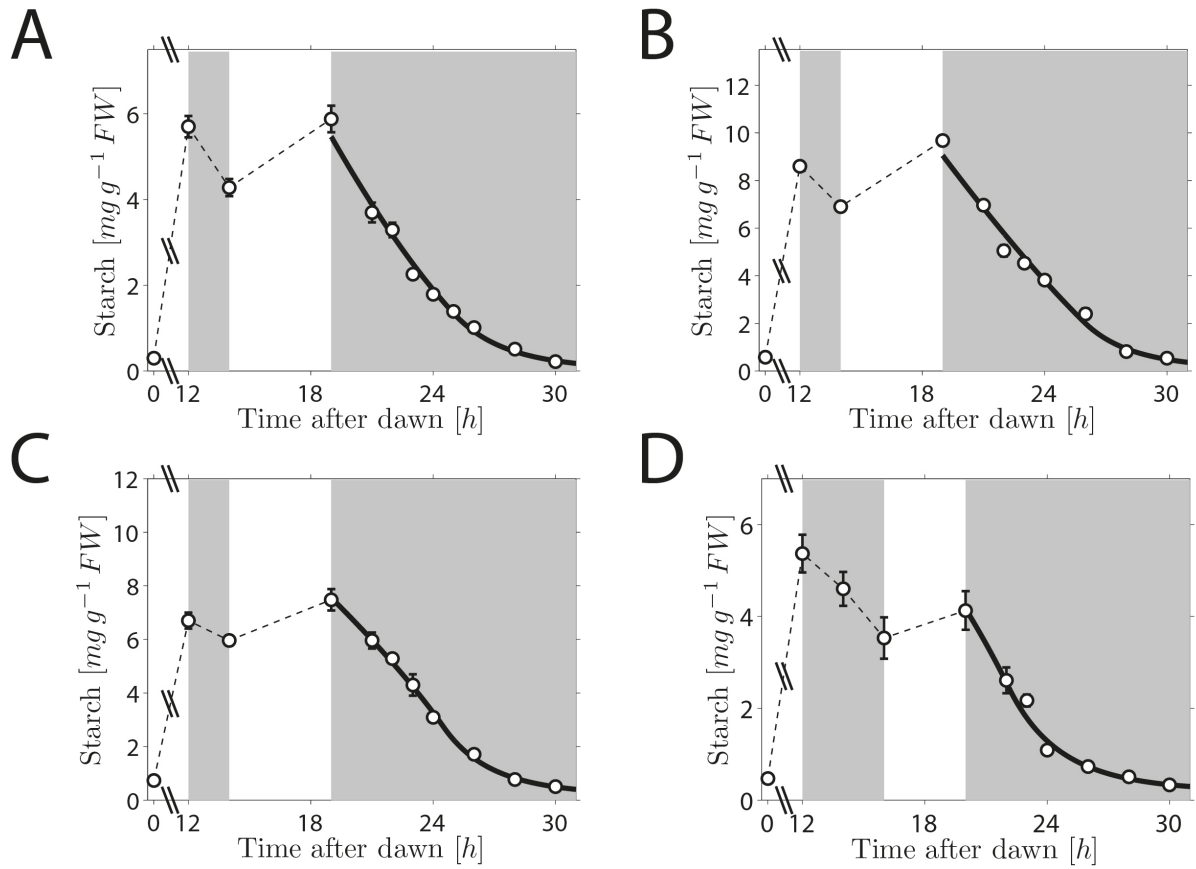

**Figure 3 – Figure Supplement 2.**

**Best fits (full lines) of the second division model to starch content data from experiments incorporating night-time light period.** Error bars are standard error of the mean throughout.

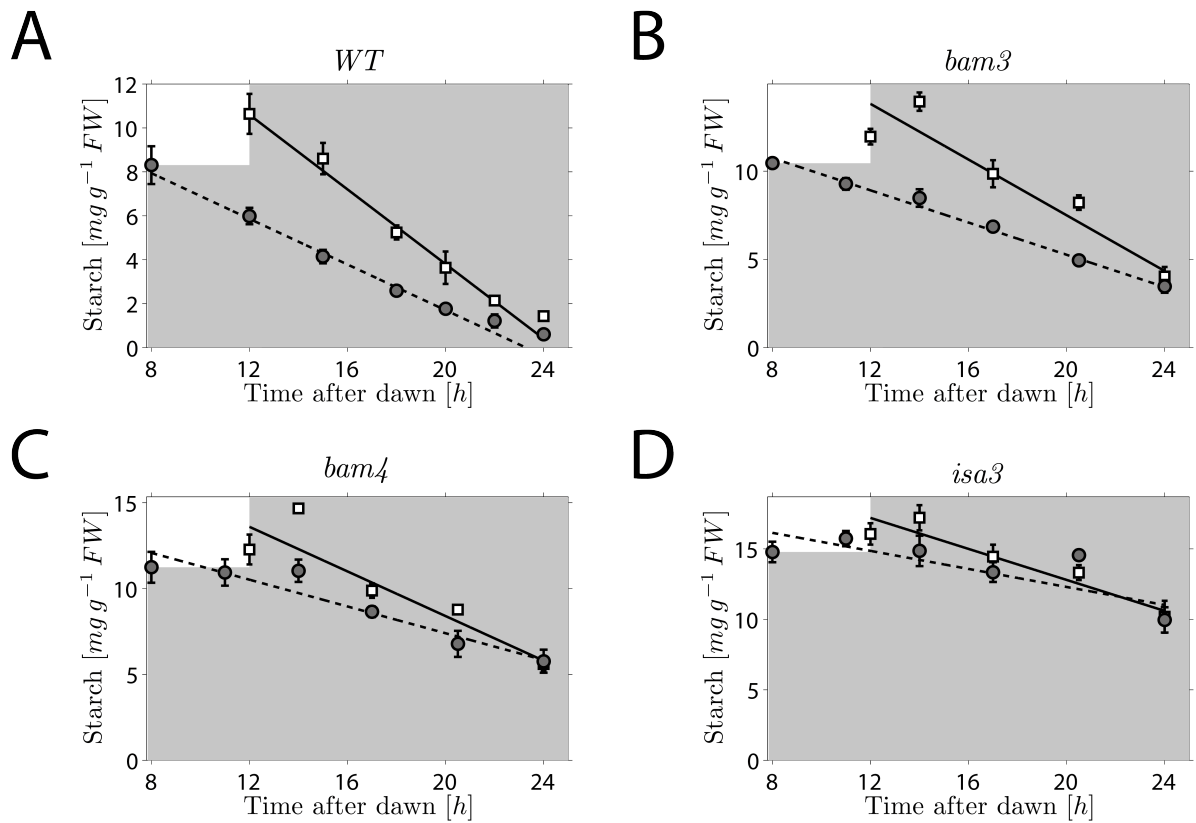

**Figure 4 – Figure Supplement 1.**

**Starch content levels during unexpectedly early night in wild-type, *bam3, bam4, isa3* mutant plants.** Starch content in wild-type (WT), *bam3, bam4, isa3* mutant Arabidopsis plants grown in 12 h light, 12 h dark cycles then subject to unexpected early (8 h, circles) or normal (12 h, squares) onset of darkness (n=6 individual rosettes for WT, n=5 for mutants; the WT dataset analyzed here is the one already shown in Figure 1A). The continuous and dashed lines are linear fits to the normal and early night datasets respectively. Error bars are standard error of the mean throughout.